\documentstyle[art12]{article}
\begin{document}
\begin{center}
{\large\bf
Diffractive charm photoproduction
\\
at HERA  $ep$-collider}

\vspace{2cm}
A.V.Berezhnoy\footnote{ Skobeltsyn Institute for Nuclear Research
Of Moscow State University, Moscow, Russia.},
V.V.Kiselev\footnote{Institute for High Energy
Physics, Protvino, Russia.}\\
$\rm I.A.Korzhavina^1$, $\rm A.K.Likhoded^2$
\end{center}

\vspace{2cm}
\begin{center}
\parbox{15cm}{
\small
\small
The cross section of the $D^*$-meson diffractive photoproduction
at the HERA collider has been calculated in the
framework of perturbatively motivated model for the different
kinematic regions. The camparison between the different Pomeron
models has been performed.
}
\end{center}

\newpage

From the experimental data,  collected at HERA $ep$-collider \cite{ZEUS},
one can conclude
that the charm production model proposed in article \cite{BKL} (below
denoted as BKL), allows to describe  inclusive
photoproduction and deep inelastic production of $D^{*\pm}(2010)$-mesons
(below denoted as $D^*$-mesons),
as well as inclusive photoproduction $D_s$-mesons,
with a good accuracy.
 
Recently new data on $D^*$-meson diffractive
photoproduction have been presented
by the ZEUS collaboration \cite{DIFZEUS}.
In this connection we try to describe these new data in the framework of
the BKL model.
 
Let us remind the general features of this model.
In the BKL approach one needs to produce perturbatively
$c$- and $\bar d$-quarks
which softly form $D^*$-meson. These perturbatively produced $c$- and
$\bar d$-quark are valence ones for the meson.
The soft hadronization process of the $c\bar d$-pair color singlet
state is described  by the average value of the operator:
\begin{equation}
\langle O_{(1)} \rangle =
\frac{1}{12 M_{D^*}} \left(-g^{\mu\nu}+\frac {p^\mu_{D^*}
p^\nu_{D^*}}{M_{D^*}^2}\right)\;
\langle D^*(p_{D^*})|(\bar c \gamma_\mu d) (\bar d \gamma_\nu
c)|D^*(p_{D^*})\rangle,
\end{equation}
where $p_{D^*}$ is  $D^*$-meson momentum, and $M_{D^*}$ is the  meson mass.
In the framework of the nonrelativistic potential model the
average value of the operator correspond with the squared wave function
in the origin:
$\langle O_{(1)}\rangle|_{NR} =|\Psi(0)|^2$.  The hadronization of the
color octet
state is described by the average value of the analogous
operator:
\begin{equation}
\langle O_{(8)} \rangle =
\frac{1}{8 M_{D^*}} \left(-g^{\mu\nu}+\frac{p^\mu_{D^*}
p^\nu_{D^*}}{M^2_{D^*}}\right)\;
\langle D^*(p_{D^*})|(\bar c \gamma_\mu \lambda^a d)
(\bar d \gamma_\nu \lambda^b
c)|D^*(p_{D^*})\rangle\; \frac{\delta^{ab}}{8}.
\end{equation}

It is worth to mention that the BKL model is based on the partonic concept
of  the hadronic structure. Indeed, in the framework of the
partonic model for the system of the infinite momentum the valence quark
structure functions are determined as follows:
\begin{equation}
\displaystyle
f_c^v(x,p_T) = f_c(x,p_T)-f_{\bar c}(x,p_T),\atop
\displaystyle
f_{\bar d}^v(x,p_T) = f_{\bar d}(x,p_T)-f_{d}(x,p_T),
\end{equation}
where  $p_T$ is a transverse momentum of a parton inside  the
hadron, and  $x$  is the fraction of the hadronic momentum taken away by
the parton. The  average  of  momentum fractions taken away by the valence
quarks  are described by the following equations:
\begin{equation}
\displaystyle
\langle x^v_c\rangle =\int d^2 p_T dx\; x\cdot f_c^v(x,p_T) \approx
\frac{m_c}{M_{D^*}},\atop
\displaystyle
\langle x^v_{\bar d}\rangle =\int d^2 p_T dx\; x\cdot f_{\bar
d}^v(x,p_T) \approx  \frac{\bar \Lambda}{M_{D^*}},
\end{equation}
where  $\langle x^v_c\rangle+\langle x^v_{\bar d}\rangle \approx 1$, and
$\bar \Lambda$ is the energy of the quark coupling into the meson.
In the framework of the BKL model we neglect
the quark velocity difference and assume  quark velocities to be
equal to each
other: $v_c = v_{\bar d}$.
The effective mass of the light quark $m_d$ in such approach plays
the role of the infrared cut. We take this mass equal to $\bar \Lambda$.
Thus we obtain the following equations:
\begin{equation}
\displaystyle
\langle x^v_c\rangle = x^v_c = \frac{m_c}{M_{D^*}},\atop
\displaystyle
\langle x^v_{\bar d}\rangle =x^v_{\bar d} = \frac{m_d}{M_{D^*}}.
\end{equation}

Now it is clear that the BKL model is an extention of the parton model
for the case of final hadrons in the framework of the valence quark
approximation.

This fact distinguishes  the  BKL  approach from perturbative
calculations \cite{Frixione, Kniehl,Cacciari}
based on the fragmentation model of the hadronisation.
In the framework of the fragmentation model of the hadronization
it is  supposed that the perturbatively produced single $c$-quark
becomes a meson at large distances. This meson
gets a fraction $z$ of the $c$-quark transverse momentum $k_T$
with the probability  determined by the fragmentation function
$D_{c \to D^*}(z,\mu)$:
\begin{equation}
\frac{d^2\sigma_{D^*}}{dzdp_T^{D^*}}=
\left.\frac{d\hat \sigma_{c \bar c}(k_T,\mu)}{dk_T}
\right|_{k_T=\frac{p_T}{z}}\cdot \frac{D_{c \to D^*}(z,\mu)}{z},
\label{fact}
\end{equation}
where  $D_{c \to D^*}(z,\mu)$ is  normalized
to the probability of $c$-quark to become a  $D^*$-meson  $w(c  \to
D^*)$, measured in  the  $e^+e^-$-annihilation  \cite{OPAL}  ($w(c\to
D^*)=0.22\pm 0.014\pm 0.014$).
$\mu$ is the scale at which the
perturbative partonic cross section of the $c\bar c$-pair production
 $d\hat \sigma_{c\bar c}/dk_{T}$ is calculated.

It is clear that in the fragmentation approach one can not take into
account the possibility for the
$c$-quark to hadronize via the
interaction  with the quark sea of the initial hadron
(recombination mechanism). That is why one needs to  account
this opportunity for $c$-quark to become a $D^*$-meson in the frame
work of some additional model.

In the framework of the BKL approach
both fragmentation and recombination mechanisms  are accounted
naturally in the calculations.
It is worth to mention that both the  fragmetation
mechanism and the recombination one are desctibed by the same set of
the diagrams.

The fragmentation mechanism dominates at the large transverse momentum
of the $D^*$-meson in accordance with the factorization theorem.
The main contribution at small transverse momenta is due to the
recombination mechanism, i.e.  due  to  the  fusion of
$c$-quark and a light quark from the sea of the initial hadron.
It is worth to mention that the recombination contribution mechanism
corresponds with higher twist cotribution
to the transverse momentum distribution.

As it was shown in many articles, devoted to the fragmentation  model,
the particular form of the fragmentation function affects
calculation results insignificantly.
In the majority of the articles
the Peterson \cite{Peterson}
fragmentation function is used:
\begin{equation}
D(z)=N\frac{1}{z(1-\frac{1}{z}-\frac{\epsilon}{(1-z)})^2},
\label{Pet}
\end{equation}
where  $N$ is normalization factor,
and $\epsilon$ is a free phenomenological parameter dependent
on the scale $\mu$.
The results would not change crucially
if  the parameterization of
Kartvelishvili-Likhoded-Petrov \cite{KLP} is used:
\begin{equation}
D(z)=Nz^{-\alpha_c}(1-z)^{\gamma -\alpha_d},
\label{Kart}
\end{equation}
where  $\alpha_c=-3$, $\alpha_d=1/2$ and $\gamma=3/2$.

In the BKL model at large tranverse momenta ($p_T^{D^*}>20\ {\rm Ē'}$)
the $D^*$-meson cross-section can be expressed by formula (\ref{fact}),
if the following pertubatively motivated form for fragmentation
function is used \cite{Frag}:
\begin{eqnarray}
\displaystyle
D_{c \to D^*}(z)&=&
\frac{8\alpha_s^2\langle O^{eff} \rangle}{27 m_d^3}
\frac{rz(1-z)^2}{(1-(1-r)z)^6}
[2-2(3-2r)z+3(3-2r+4r^2)z^2-
\nonumber
\\
&&
2(1-r)(4-r+2r^2)z^3+
(1-r)^2(3-2r+2r^2)z^4],
\end{eqnarray}
where   $r=m_d/(m_d+m_c)$ and
\begin{equation}
\langle O^{eff}\rangle=\langle O_{(1)}\rangle + \frac{1}{8}
\langle O_{(8)} \rangle.
\label{ratio}
\end{equation}

Meson  production in $e^+e^-$-annihilation is caused
by fragmentation mechanism only. This fact  allows to define
 $\langle O^{eff}\rangle$ using $w(c \to D^*)$ and quark masses:
\begin{equation}
w(c \to D^*)=\int_0^1 D_{c\to D^*}(z)dz=
\frac{\alpha_s^2(\mu)\langle O_{(1)}(\mu)\rangle
}{m_d^3} \cdot I(r),
\label{prob}
\end{equation}
where   $I(r)$ is determined in \cite{BKL}.
The value of $w(c \to D^*)$ is known from the experiment.

So one can determine the value of $\langle O^{eff}\rangle$  for
fixed values of $m_d$, $m_c$ and $\mu$. Assuming
\begin{equation}
\begin{array}{rcl}
\mu &=& m_{D^*},\\
m_d &=& 0.3 \ {\rm GeV},\\
m_c &=& 1.5 \ {\rm GeV} \ {\rm and}\\
w(c\to D^*) &=& 0.22.\\
\end{array}
\label{value}
\end{equation}
one obtains
$$\langle O^{eff}(m_{D^*}) \rangle  = 0.25 \ {\rm GeV}^{3}.$$

The best description of the experimental data on the charm photoproduction
and the deep inelastic charm production at HERA can be achieved
if $\langle O_{(8)} \rangle / \langle O_{(1)}\rangle =1.3$.

The mentioned values of the parameters were taken
both for the presented calculations and
for the calculations of the $D^*$-meson nondiffractive
production cross section \cite{BKL}.

BKL model can be used for all values of $D^*$-meson
transverse momentum in
the contrast to the fragmentation model which is valid only for large
transverse momentum (our estimation shows that fragmentation model
can be used for $p_T^{D^*}>20$~GeV).

We use the parameter values (\ref{value}) to calculate
the cross-section of the $D^*$-meson diffractive production
in the frame work of the BKL model.

We choose one of the known form of the Pomeron flux
parametrization \cite{flux}:
\begin{equation}
f_{{ I\!\!P}/p}(x_{ I\!\!P},t)=\frac{1}{2} \frac{1}{2.3}
\frac{1}{x_{ I\!\!P}}\left[ 6.38e^{-8|t|}+0.424e^{-3|t|}\right],
\end{equation}
where  $t$ is the squared transfer momentum
in the proton vertex, and  $x_{ I\!\!P}$ is the momentum fraction
of the proton carried away by the Pomeron.  We neglect the amplitude
dependence on $t$ in the calculations.

We suppose that the Pomeron consists of the gluons only  and
used two types of the gluonic distribution inside the Pomeron $G(\beta)$:

\begin{equation}
\beta G(\beta)=\left\{ {6\beta(1-\beta)\ \ - {\rm  ``hard"\  Pomeron}};
\atop {6(1-\beta)^5\ \ - {\rm ``soft"\ Pomeron,}}\right.
\end{equation}
where  $\beta$ is  the fraction of the Pomeron momentum
carried off by a gluon.

In Fig.1 the calculation of the differential distributions
of the $D^*$-meson production has been performed for
the kinematic region investigated  by ZEUS Collaboration\cite{DIFZEUS}:
$130<W<280\ {\rm GeV}$, $Q^2<1 \ {\rm GeV^2}$,
 $p_{T}^{D^*}>2 \ {\rm GeV}$, $|\eta^{D^*}|<1.5$,
$0.001<x_{ I\!\!P}<0.018$, where  $W$ is the invariant
mass of the photon-proton system,  $Q^2$ is the photon virtuality
and  $\eta^{D^*}$ is the pseudorapidity of the $D^*$-meson.
The value of  $\eta^{D^*}$ is determined by the angle $\theta$ between
the initial proton direction and the $D^*$-meson
in the laboratory system as follows:
$\eta^{D^*}=-{\rm ln}({\rm tg}\frac{\theta}{2})$.
 
The cross sections calculated for this kinematic region 
in the frame work of BKL model  
have the following numerical values:
$$
\sigma^{\rm BKL}=\left\{ { 0.77\pm 0.02\ \ - {\rm  ``hard"\  Pomeron}};
\atop { 0.56\pm 0.03\ \ - {\rm ``soft"\ Pomeron.}}\right.
$$
 
From the transverse momentum distributions one can conclude, that
the calculation with the soft Pomeron predicts more rapid decreasing
of the cross section with increasing transverse momentum than
the calculation with the hard Pomeron.

From the distribution in $D^*$ pseudorapidity one can see
that the model of the hard Pomeron predicts
the cross section values of the  $D^*$ meson production
in the forward direction essentially larger than
the ones predicted by the soft Pomeron model.
The predictions of these models for the $D^*$-meson
backward  production in the considered kinematic region
are practically the same.

The hard Pomeron model predicts the distribution maximum at small $M_x$
values.
The soft Pomeron model calculation yields the distribution  maximum
at $M_X$ over 20~GeV.

In the considered range of $x_{I\!\!P}$ the differential cross sections
calculated with different models of Pomeron, show
quite different behaviour: hard Pomeron decreases with
 $x_{I\!\!P}$ while soft Pomeron increases instead.

The ZEUS Collaboration plans to continue the analysis of the diffractive
photoproduction of the $D^*$ meson. That is why we have calculated the
total and the differential cross sections
for $0.001<x_{I\!\!P}<0.2$ (Fig.~2):
$$
\sigma^{\rm BKL}=\left\{ { 1.64\pm 0.02\ \ - {\rm  ``hard"\  Pomeron}};
\atop { 5.45\pm 0.02\ \ - {\rm ``soft"\ Pomeron.}}\right.
$$

and for the $0.001<x_{I\!\!P}<0.1$ (Fig.~3):
$$
\sigma^{\rm BKL}=\left\{ { 1.51\pm 0.03\ \ - {\rm  ``hard"\  Pomeron}};
\atop { 3.56\pm 0.03\ \ - {\rm ``soft"\ Pomeron.}}\right.
$$

It is evident that for these ranges of  $x_{I\!\!P}$
the difference between the soft Pomeron model and the
hard Pomeron model is more essential than for the $0.001<x_{I\!\!P}<0.018$.
The soft Pomeron model leads to the larger value of the cross
section than the hard Pomeron model.

These researches have been partially supported by RFBR Grants 00-15-96645
and 99-02-16558.

\newpage\vspace*{-3cm}
\includegraphics{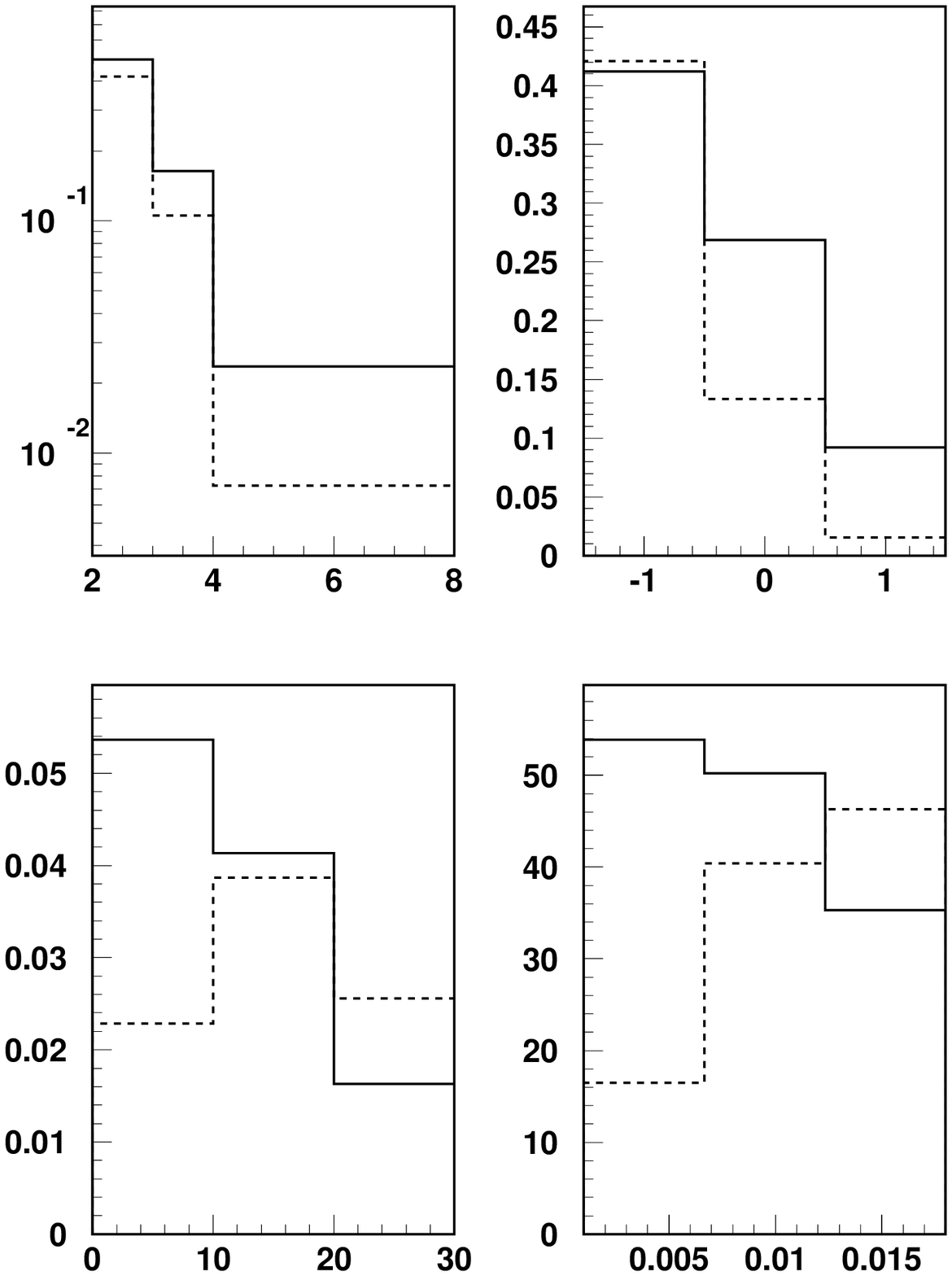}
\begin{picture}(450,450)\put(15,405){$d\sigma/d p_T^{D^*}$, nb/GeV}
\put(15,120){$d\sigma/dM_x$, nb/GeV}
\put(250,405){$d\sigma/d\eta^{D^*}$, nb}
\put(250,120){$d\sigma/d x_{ I\!\!P}$, nb}
 \put(420,135){$\eta^{D^*}$}
\put(170,135){$p_T^{D^*}$, GeV}
\put(170,-150){$M_X$, GeV}
\put(145,330){\it a)}
\put(145,45){\it c)}
\put(420,-150){$x_{ I\!\!P}$}
\put(380,330){\it b)}
\put(380,45){\it d)}
\put(-10,-180){Fig.~1. \parbox[t]{14cm}{
The BKL  model  predictions  for  the differential cross
sections of the $D^*$ diffractive photoproduction  at HERA:
a) $p_{T}^{D^*}$;
b) $\eta^{D^*}$;
c) $M_X$;
d) $x_{ I\!\!P}$ are calculated in the kinematic region:
$130<W<280\ {\rm GeV}$, $Q^2<1 \ {\rm GeV^2}$,
   $p_{T}^{D^*}>2 \ {\rm GeV}$, $|\eta^{D^*}|<1.5$,
  $0.001<x_{ I\!\!P}<0.018$.
The solid histograms stand for the
calculations with the hard Pomeron model, and the dashed ones stand
for the calculations with the soft Pomeron one.
}}
\end{picture}

\newpage\vspace*{-3cm}
\includegraphics{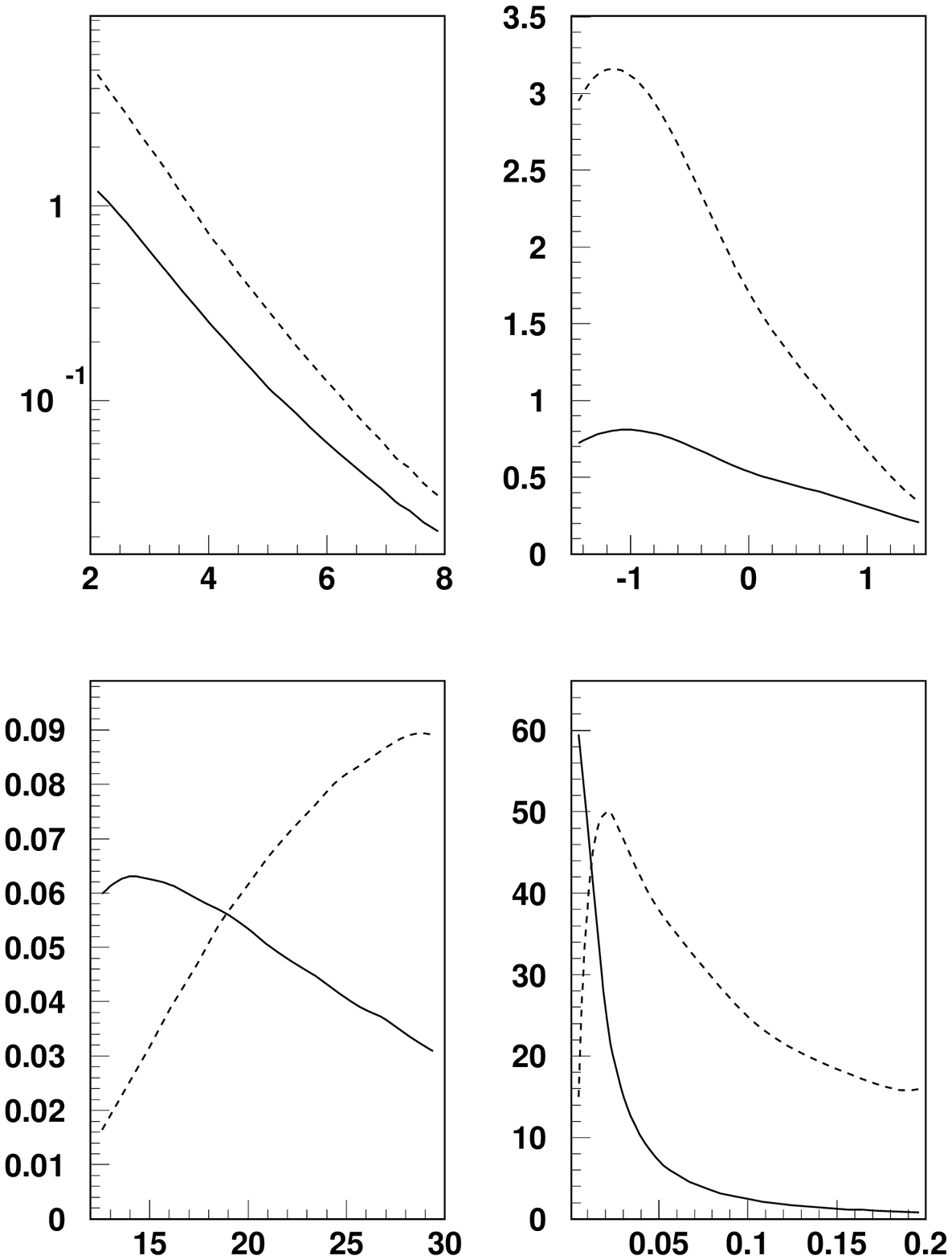}
\begin{picture}(450,450)\put(15,405){$d\sigma/d p_T^{D^*}$, nb/GeV}
\put(15,120){$d\sigma/dM_x$, nb/GeV}
\put(250,405){$d\sigma/d\eta^{D^*}$, nb}
\put(250,120){$d\sigma/d x_{ I\!\!P}$, nb}
 \put(420,135){$\eta^{D^*}$}
\put(170,135){$p_T^{D^*}$, GeV}
\put(170,-150){$M_X$, GeV}
\put(145,330){\it a)}
\put(145,45){\it c)}
\put(420,-150){$x_{ I\!\!P}$}
\put(380,330){\it b)}
\put(380,45){\it d)}
\put(-10,-180){Fig.~2. \parbox[t]{14cm}{
The BKL model predictions  for the differential cross section
of the $D^*$ diffractive photoproduction  at HERA:
a) $p_{T}^{D^*}$;
b) $\eta^{D^*}$;
c) $M_X$;
d) $x_{ I\!\!P}$ are calculated in the kinematic region:
$130<W<280\ {\rm GeV}$, $Q^2<1 \ {\rm GeV^2}$,
   $p_{T}^{D^*}>2 \ {\rm GeV}$, $|\eta^{D^*}|<1.5$,
  $0.001<x_{ I\!\!P}<0.2$.
The solid curves stand for the
calculations with the hard Pomeron model,
and the dashed ones stand
for the calculations with the soft Pomeron model.
}}
\end{picture}
\newpage\vspace*{-3cm}
\includegraphics{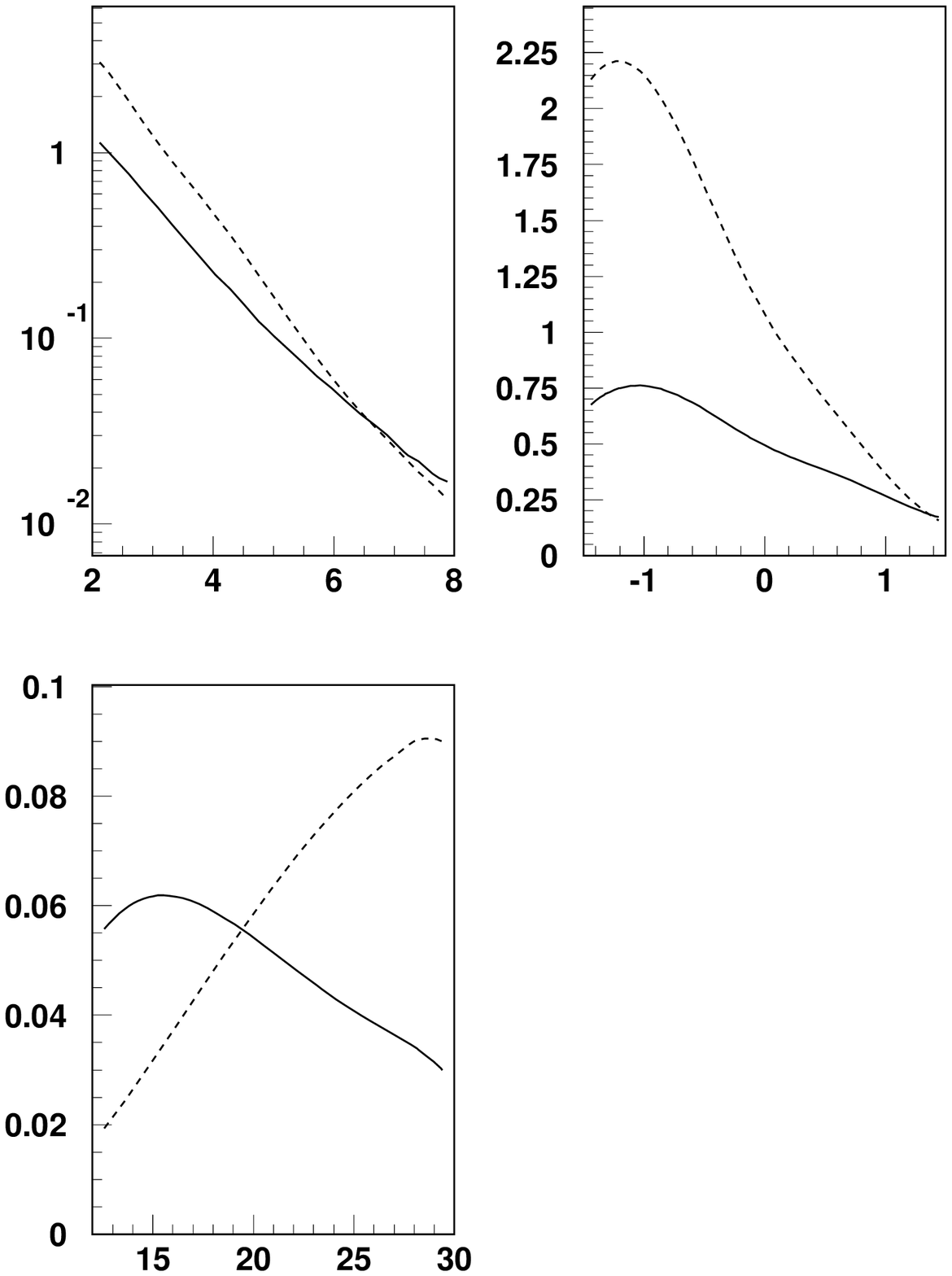}
\begin{picture}(450,450)\put(15,405){$d\sigma/d p_T^{D^*}$, nb/GeV}
\put(15,120){$d\sigma/dM_x$, nb/GeV}
\put(250,405){$d\sigma/d\eta^{D^*}$, nb}
 \put(420,135){$\eta^{D^*}$}
\put(170,135){$p_T^{D^*}$, GeV}
\put(170,-150){$M_X$, GeV}
\put(145,330){\it a)}
\put(145,45){\it c)}
\put(380,330){\it b)}
\put(-10,-180){Fig.~3. \parbox[t]{14cm}{
The BKL model predictions  for the differential cross section
of the $D^*$ diffractive photoproduction  at HERA:
a) $p_{T}^{D^*}$;
b) $\eta^{D^*}$;
c) $M_X$ are calculated in the kinematic region:
$130<W<280\ {\rm GeV}$, $Q^2<1 \ {\rm GeV^2}$,
   $p_{T}^{D^*}>2 \ {\rm GeV}$, $|\eta^{D^*}|<1.5$,
  $0.001<x_{ I\!\!P}<0.1$.
The solid curves stand for the
calculations with the hard Pomeron model,
and the dashed ones stand
for the calculations with the soft Pomeron model.
}}
\end{picture}


\begin{thebibliography}{35}
\bibitem{ZEUS}
ZEUS Collab. ( J. Breitweg {\it et al.}), Eur.Phys.J. C {\bf 6}, 67(1999).
\bibitem{ZEUS_Ds}
ZEUS Collab.  ( J.  Breitweg {\it et al.}),  Phys.Lett.  B {\bf  481},
213(2000).
\bibitem{BKL}
A.V. Berezhnoy, V.V. Kiselev, A.K. Likhoded,
Phys.Rev. D {\bf 62}, 074013(2000).
\bibitem{DIFZEUS}
I.A. Korzhavina, representing the ZEUS Collaboration,
Talk given at the Russian Academy of Science (RAS)
Nuclear Physics 2000 Conference, Moscow, Russia,
 November 27-December 2, 2000. To be published in Rus. Nucl. Phys.
(Yad.Fiz.), Proceedings of RAS Nucl. Phys. 2000 Conf.
\bibitem{Frixione}
S. Frixione,  M.L.  Mangano,  P.  Nason, G. Ridolfi, Phys.Lett. B {\bf
348}, 633(1995);
S. Frixione, P. Nason, G. Ridolfi, Nucl.Phys. B {\bf 454}, 3(1995).
\bibitem{Kniehl}
B.A. Kniehl, G. Kramer, M. Spira, Z.Phys. C {\bf 76}, 689(1997);
J. Binnewies, B.A. Kniehl, G. Kramer, Z.Phys. C {\bf 76}, 677(1997);\\
J. Binnewies, B.A. Kniehl, G. Kramer, Phys.Rev.  D {\bf 58}, 014014(1998).
\bibitem{Cacciari}
M. Cacciari, M. Greco, Z.Phys. C {\bf 69}, 459(1996);
M. Cacciari, M. Greco, S. Rolli, A. Tanzini, Phys.Rev. D {\bf 55}
2736(1997);
M. Cacciari, M. Greco, Phys.Rev. D {\bf 55}, 7134(1997).
\bibitem{OPAL} 
OPAL Collab. (K. Akerstaff {\it et al.}), Eur.Phys.J. C {\bf 1}, 439(1998).
\bibitem{Peterson} 
C. Peterson  {\it et al.}, Phys.Rev.  D {\bf 27}, 105(1983).
\bibitem{KLP}
V.G. Kartvelishvili, A.K. Likhoded, V.A. Petrov, Phys.Lett. B {\bf 78},
615(1978).
\bibitem{Frag} 
 C.-M. Chang, Y.-Q. Chen, Phys.Rev. D {\bf 46}, 3845(1992),
D {\bf 50}, 6013(E)(1994);
E. Braaten, K. Cheung, T.C. Yuan, Phys.Rev. D {\bf 48} 4230(1993);
V.V. Kiselev, A.K. Likhoded, M.V. Shevlyagin, Z.Phys. C {\bf 63}, 77(1994);
T.C. Yuan, Phys.Rev. D {\rm 50}, 5664(1994);
K. Cheung, T.C. Yuan, Phys.Rev. D {\bf 53}, 3591(1996).
\bibitem{flux}
P. Bruni, G. Ingelman, Phys.Lett. B {\bf 331}, 317(1993);
P. Bruni, G. Ingelman, Diffractive hard scattering at $ep$ and $p\bar p$
colliders, in Proc. of the EPS International High Energy Physics Conference,
edited by J.Carr, M.Perrottet
(Editions Frontieres, Marseille, France, 22-28 Jul, 1993), DESY 93-187;
P. Bruni, G. Ingelman, A. Solano, Diffractively produced hadronic
final states and the pomeron structure, in Proc. of the Workshop on Physics
at HERA, Vol.1, edited by W.Buchm\"uller, G.Ingelman (DESY, Hamburg, 1991),
Vol. 311, p.363.
\end{thebibliography}
\end{document}